\documentstyle[preprint,aps]{revtex}
\tightenlines 
\title{ Quantization and Corrections of   Adiabatic
Particle Transport in 
a Periodic  Ratchet Potential}
\author{Yu Shi$^{1}$\cite{s} and Qian Niu$^{2,3}$\cite{n}} 
\address{(1) Theory of Condensed Matter, Cavendish Laboratory, 
University of Cambridge, 
 Cambridge CB3 0HE, United  Kingdom\\
(2) Department of Physics, University of Texas, Austin, TX 78712\\
(3) ICQS, Chinese Academy of Sciences, Beijing 100080, China}
\date{Europhys. Lett. {\bf 59}, 324 (2002) }
\begin{document}
\draft
\maketitle

\begin{abstract}

We study the transport of 
an overdamped particle  adiabatically driven by an 
asymmetric potential which is periodic in both space and time.
We develop an   adiabatic  perturbation theory
after transforming the Fokker-Planck equation   into 
a  time-dependent hermitian problem, and reveal  an   analogy
with  quantum adiabatic particle transport.
An analytical expression is obtained  for   the      
ensemble average of the particle velocity in terms of the Berry phase of 
the Bloch states. Its time average is 
shown to be  quantized as a Chern number
in the  deterministic or tight-binding limit, with exponentially
small corrections. In the opposite limit, where the thermal energy dominates
the ratchet potential, 
a formula for the average velocity  is also obtained, showing
a second order dependence on the potential.

\end{abstract}

\pacs{PACS numbers: 05.40.-a,  02.40.-k,  72.10.-d, 87.10.+e}
 

The so-called   ratchet  models
have attracted  much  interest in recent years~\cite{reimann,prost,magnasco}.  
In  a typical   ratchet model,  a  driving force with  
time-correlated fluctuation, in addition to a random
force,    is exerted 
on an overdamped particle.  A  macroscopic 
flow develops if the parity symmetry is broken in the driving potential. 
This behavior is in striking 
contrast to the equilibrium situation in which
the second law of thermodynamics 
forbids a non-vanishing macroscopic flow, irrespective of the microscopic 
details~\cite{feynman}. 
Here we consider  an adiabatic  ratchet 
potential with  both  spatial and temporal  periodicities,
which  has received some attention recently~\cite{bartussek,parrondo}.
We take an  analytical approach by 
transforming the Fokker-Planck equation into a time-dependent
 hermitian problem.  The ensemble average of the
particle velocity is calculated by
using the standard adiabatic perturbation theory, whose validity condition is 
also explicitly specified for the present system. The particle transport 
in a time cycle is found in terms of the Berry phase of the Bloch waves.  
In  the   deterministic or 
tight-binding limit, in which the potential dominates the thermal 
energy, the particle transport is shown to be quantized as a Chern 
number~\cite{chern},  consistent with the previous  numerical 
result~\cite{bartussek}, and in analogy  with  quantum adiabatic particle
transport first studied by Thouless and Niu~\cite{thouless,nt,niu}.
The correction to the quantization  is found to be  exponentially 
small based on an analysis using the Wannier functions. We also elaborate 
the opposite limiting case, where the thermal energy dominates the
potential.

Consider an overdamped particle moving in one-dimension, subject to a  driving force $F(x,t)=-\partial_x V(x,t)$  and
a random force $\nu(t)$. It obeys a Langevin equation,  
\begin{eqnarray*}
\gamma\frac{dx}{dt} = \frac{1}{\mu}F(x,t)+\nu(t),
\end{eqnarray*}
where $\gamma$ is the friction constant and $\mu$ is the mass.  
We assume that $\nu(t)$ is  of zero average and  $\delta$-correlated, i.e. 
$\langle\nu(t)\nu(t')\rangle=2(\gamma k_{B}T/\mu)\delta(t-t')$, where 
$k_B$ is the Boltzmann constant, 
$T$ is  the temperature, and $\langle\cdots\rangle$ denotes  the 
ensemble average. The potential  $V(x,t)$ is 
  periodic in both space and time, with periods $a$ and 
$\tau$ respectively. Furthermore, the potential is assumed to be asymmetric, 
i.e. $V(x)\neq V(-x)$, as shown in  Fig. 1. 

The corresponding  Fokker-Planck equation for the probability density
$\rho(x,t)$ is
\begin{equation}
-\partial_t \rho(x,t) = D{\cal O}\rho(x,t), 
\label{FP}
\end{equation}
where $D=k_BT/\mu\gamma$ is the diffusion constant, and
${\cal O}=-\partial_{x}^{2}+\partial_x\cdot F/k_BT$.
The Fokker-Planck equation (\ref{FP}) can also  be written 
as a continuity equation $\partial_t \rho+\partial_x j=0$, 
where the probability 
current is  $j(x,t)\equiv D\hat{j}\rho(x,t)$, with 
$\hat{j} =-\partial_{x}+F/k_BT$. 

The operator ${\cal O}$ is non-hermitian, but there is 
a similarity transformation~\cite{zinn} leading   to  a hermitian 
operator, to be called the effective hamiltonian,  
\begin{equation}
{\cal H}=e^{\Gamma}{\cal O}e^{-\Gamma}
=\hat{p}^2+U,
\label{tr}
\end{equation}
where 
$\Gamma=V/2k_BT$
is  the  potential normalized by  the thermal energy, 
 $U=(\partial_x\Gamma)^2-\partial_x^2\Gamma$ is the  effective potential,
and $\hat{p}=-i\partial_x$.
Note that ${\cal H}$ has the dimension of inverse length squared.
Since ${\cal H}$ is hermitian, its  eigenfunctions   comprise 
a complete orthonormal basis.  These are the Bloch waves, 
$\psi_{nk}$, owing to the periodicity of the potential, where 
$n$ is the discrete band index, and $k$ is the Bloch vector.
It is also interesting and useful to note that the ground state
eigenvalue is zero, and the wave function can be explicitly found as
$$\psi_{0}=\frac{1}{\sqrt{Z}}e^{-\Gamma}, $$
where  $Z=\int e^{-2\Gamma} dx$ is a normalization factor ensuring
$\langle \psi_{0}|\psi_{0}\rangle=\int \psi_{0}(x)^*\psi_{0}(x) dx =1$.

The Fokker-Planck equation is then transformed into
\begin{equation}
-\partial_t\psi(x,t)=
(D{\cal H}-\partial_t\Gamma-\partial_t\ln\sqrt{Z})\psi(x,t),
\label{hp}
\end{equation}
where $\psi=\rho e^\Gamma \sqrt Z$ is the transformed variable for the
probability  density,
 and the  factor $\sqrt Z$ is added  for  convenience. 
If $\psi=\psi_{0}$, then the probability density is 
the stationary  Boltzmann distribution
$\rho = e^{-2\Gamma}/Z$, giving $j=0$.  
If $V$ were time-independent, the above equation would reduce to 
$-\partial_t \psi = D{\cal H}\psi$,  whose ground state solution  
$\psi=\psi_{0}$ corresponds to the equilibrium state.

The wave function $\psi$ is generally 
a superposition of different Bloch states.  However, only those
with k=0 are  important if we are only interested in 
the ensemble average of the particle velocity 
$\langle \frac{dx}{dt} \rangle$.  This  
is equal to the spatial integral of the
probability current,
\begin{equation}
J(t)=D\int \hat{j} \rho dx
=D\int \psi_{0}(-\partial_x-\partial_x\Gamma)\psi dx
 = -2D \int \psi_{0}\partial_x \psi (x,t) dx,
\label{jj}
\end{equation}
where,  in the expansion of $\psi(x,t)$,  only 
$\psi_{nk}(x,t)$ with $k=0$ contribute. 
This result does not require the necessity of adiabaticity.

We now apply the adiabatic perturbation theory ~\cite{kato}, supposing
that the ratchet potential changes slowly. 
We first make the expansion
\begin{equation}
\psi(x,t)=\sum_n   c_{n}(t)\psi_{n}(x,t)e^{-D\int_0^t E_{0}(t')dt'},
\label{ex1}
\end{equation}
where $k=0$ is assumed in light of the discussions above.  For clarity,
we have retained $E_0$ although it is zero.  The coefficient for the ground state is constrained to $c_{0}(t)\equiv 1$ as imposed by 
the normalization of the density $\rho(x,t)$.  The coefficients for the higher bands satisfy
\begin{equation}
D(E_{0}-E_{n})c_{n}
+\sum_{n'}c_{n'}\langle \psi_{n}|\partial_t\Gamma+\partial_t\ln\sqrt{Z}
-\partial_{t}|\psi_{n'}\rangle=\partial_t c_{n}.
\end{equation} 
 The terms in the summation are small if the potential changes slowly 
and if we choose the Bloch waves with $k=0$ to be real 
such that $<\psi_{n}|\partial_t|\psi_{n}>=0$.  
We also assume that the system is initially in the ground state (stationary
state). Thus we can ignore all but the $n'=0$ term in the summation, 
obtaining, to first order in the rate of change of the potential,
\begin{equation}
c_{n\neq 0}(t)= \frac{2\langle\psi_{n}|\partial_t\psi_{0}\rangle}
{D[E_{0}(t)-E_{n}(t)]}, 
 \label{ca}
\end{equation}
where we have used the fact that 
$(\partial_t\Gamma+\partial_t\ln\sqrt{Z})\psi_{0}=-\partial_t\psi_{0}$.
The smallness of $c_n$ for $n>0$ means that the system remains 
close to quasi-static 
equilibrium with the instantaneous ratchet potential.  
If the system has initial excitations in the higher bands, one can show that 
they decay away exponentially at a 
rate of $D(E_n-E_0)$ and become negligible after a transient period. 

Quantitatively, the adiabatic condition is satisfied if the potential changes 
at a rate much smaller than $D$ times the eigenvalue
 gap, $\frac{1}{\tau} \ll D (E_1-E_0)$.
 We now make an order of magnitude estimate of this gap. 
 We assume that the amplitude of variation in the force is $F_0$, 
yielding the variation for $\Gamma$ as $\Gamma_0=F_0a/2k_BT$.
In the kinetic regime where $\Gamma_0<<1$, the gap at $k=0$ 
is about $(2\pi/a)^2$. 
 The adiabatic condition is just $\tau \gg \tau_D$,
where $\tau_D=a^2/D$ is 
the diffusion time over one wavelength of the ratchet. 
In the tight-binding or deterministic regime, where $\Gamma_0>>1$, 
the Bloch bands may be viewed as derived from the levels in the deep 
potential wells. 
A peculiar thing to notice is that the effective potential $U$ 
typically contains a double well structure in each wavelength of
 the ratchet,   
even if the ratchet potential has one well in each period.  The reason is that 
the dominant term $(\partial_x\Gamma)^2$ in $U$ has only half of wavelength of 
the ratchet, and thus must contain two wells in each wavelength of the ratchet.
These two wells are made inequivalent by the weaker term, 
$-\partial_x^2 \Gamma$, which has the full periodicity of the ratchet. 
 The band gap can thus be estimated by the level difference between the two wells, and is given by the amplitude of variation of the weaker term 
$\Gamma_0/a^2$.  The 
adiabatic condition in this case is  $\tau \gg  \tau_D/\Gamma_0 $,
which is more easily satisfied than the kinetic regime 
because of the large $\Gamma_0$ in the tight-binding regime. 

The average velocity induced by the adiabatic movement of the ratchet can
 then be obtained from (\ref{jj}), (\ref{ex1}) and (\ref{ca}) as 
\begin{equation}
J(t)=-4\sum_{n\neq 0}
\frac{\langle\psi_{0}|\partial_x\psi_{n}\rangle\langle
\psi_{n}|\partial_t\psi_{0}\rangle}{E_{0}-E_{n}},
\label{jjjj}
\end{equation}
This current is independent of the diffusion constant $D$, because all the 
eigenstates and eigenvalues
 of the effective Hamiltonian are independent of $D$.  
This signifies that the adiabatic flow is a geometric effect: it only 
depends on how the ratchet potential evolves in time, and is independent of 
the mobility of the particles.  The only role of the diffusion constant is 
to set the time scale for the validity of the adiabatic approximation.  The 
reader is also reminded that the distinction between the tight-binding and
 kinetic regimes, in which the adiabatic flow  will be shown to have very
 different behaviors, also has nothing to do with $D$.    

In the following, we make a further connection to the Berry phase, and show
when the adiabatic current may be quantized. We can write $J(t)={\cal J}(k=0,t)$, with 
\begin{equation}
{\cal J}(k,t)=-2\sum_{n\neq 0}
\left[ \frac{\langle\psi_{0k}|\partial_x\psi_{nk}\rangle\langle
\psi_{nk}|\partial_t\psi_{0k}\rangle}{E_{0k}-E_{nk}}
+\frac{\langle\partial_t\psi_{0k}|\psi_{nk}\rangle\langle
\partial_x\psi_{nk}|\psi_{0k}\rangle}{E_{0k}-E_{nk}}\right].
\label{jjj}
\end{equation}
This can be rewritten in terms of the periodic amplitude $u_{nk}(x,t)$ of the 
Bloch waves, and turned into a form involving the lowest band only
\begin{equation}
{\cal J}(k,t)=\frac{1}{ i}
\left[\langle \partial_{k}  u_{0k}|\partial_t u_{0k}\rangle -
\langle  \partial_t u_{0k}|\partial_{k}u_{0k}\rangle\right],
\label{cc0}
\end{equation}
following the method used in deriving 
the quantum adiabatic particle transport~\cite{thouless,nt}.
This is known as the Berry curvature of the Bloch state in the
parameter
 space of $k$ and $t$.
The time average of $J(t)$  over a
time cycle of the ratchet is therefore
\begin{equation}
\overline{J(t)}=\frac{1}{ i\tau}
\int_0^{\tau}dt
\left[\langle \partial_{k} u_{0k}|\partial_t u_{0k}\rangle -
\langle\partial_t u_{0k}|\partial_{k}u_{0k}\rangle\right]_{k=0}
=-{1\over \tau}\partial_k \theta(k)_{k=0},
\label{cc}
\end{equation}
where $\theta(k)=i\int dt \langle u_{0k}|\partial_{t} u_{0k}\rangle$
is the Berry phase for a given $k$.  In obtaining the last expression in 
the above equation, we assumed a phase gauge such that the wave function 
is periodic in time, which is possible because the eigenstate must come 
back to itself in a time cycle of the potential apart from a phase factor.

In the tight-binding or deterministic limit, the
quantity ${\cal J}(k,t)$ is insensitive to $k$, as can be seen
in  (\ref{jjj}), 
 which can be be written as 
\begin{equation}
{\cal J}(k,t)=2
\left[ \langle\partial_x \psi_{0k}|\frac{1}{E_{0k}-{\cal
H}}|
\partial_t\psi_{0k}\rangle +
\langle\partial_t \psi_{0k}|\frac{1}{E_{0k}-{\cal H}}|
\partial_x\psi_{0k}\rangle
\right].
\label{jr}
\end{equation}
We may thus 
replace the $k=0$ expression in (11) by an average over $k$, yielding 
\begin{equation}
\overline{J(t)}=\frac{a}{2\pi i\tau}
\int_0^{2\pi/a}dk\int_0^{\tau}dt
\left[\langle \partial_{k} u_{0k}|\partial_t u_{0k}\rangle -
\langle\partial_t u_{0k}|\partial_{k}u_{0k}\rangle\right].
\label{c}
\end{equation}
This is a closed surface integral of the Berry curvature,
and can only take quantized values, i.e.
$$\overline{J(t)}=N\frac{a}{\tau},$$ 
where $N$ is an integer called Chern number.

Corrections to this exact quantization, which arise in 
our replacing the k=0 expression by an average over $k$, can be shown 
to be exponentially small in $\Gamma_0$.  Consider first the somewhat trivial 
example of a sliding potential, $V(x-vt)$, which is in general  
not a ratchet. The position and time dependence of the
 Bloch states must be through the combination $x-vt$.  We can thus replace 
$\partial_t$ by $-v\partial_x$, turning (8) into an expression related to 
the ``effective mass'' of a band ~\cite{niu}. The result is that
the average particle current can be written 
in the form $v[1-2^{-1}\partial^2 E_{0k}/\partial k^2 |_{k=0}]$. 
 The first term is just the quantization with $N=1$, which means that
the particle follows the potential, 
while the second term is exponentially small in $\Gamma_0$ because 
the band width is so in the tight-binding limit. 

To establish the exponential smallness of the correction to quantization for 
the general case of the tight-binding regime, we introduce the the Wannier 
functions 
%
$\phi_n (x-la)= \frac{a}{2\pi}\int dk e^{-i k la }\psi_{nk},$
which are known to be exponentially localized and form an orthonormal basis 
for each band.  We may rewrite (\ref{cc0}), with $k=0$, as 
\begin{equation}
J(t)=\partial_t \langle x \rangle +
a\sum_l l \langle \phi_0(x,t)|\partial_t\phi_0(x-la,t)\rangle,
\label{cccc}
\end{equation}
where 
$\langle x \rangle \equiv \langle \phi_0(x,t)|x|\phi_0(x,t)\rangle$ 
is the center  position of the Wannier function~\cite{niu}. 
In deriving the above result, it is useful to recall the inverse relation
$\psi_{nk}(x,t)= \frac{1}{\sqrt M} \sum_l e^{ i k la } \phi_n (x-l a,t)$
between the Bloch and Wannier functions, where $M$ is the number of cells 
in the system.  Therefore 
\begin{equation}
\overline{J(t)}
=\frac{1}{\tau}(\langle x\rangle|_{t=\tau}
- \langle x \rangle|_{t=0}) + 
\frac{a}{\tau}\sum_l l \int_0^{\tau} dt 
\langle \phi_0(x,t)|\partial_t\phi_0(x-la,t)\rangle.
\label{accc}
\end{equation} 
$\langle x \rangle $ can only change by  an integer multiple of 
the lattice constant in a time period, so 
the first term in (\ref{accc}) is nothing but the 
quantization $Na/\tau$.
The second  term gives a correction to this quantization. 
Contribution from $l=0$ term is automatically zero.  
In the tight-binding limit, these Wannier functions 
decay exponentially away from  the bottom of each well,
 so the leading term in the correction 
comes from the nearest-neighbor term $l=1$.   Therefore, the correction is 
proportional to the overlap between the nearest-neighboring
Wannier functions, which is 
exponentially small in $\Gamma_0$.

On the other hand, 
beyond the tight-binding or deterministic limit, the quantization is
destroyed.  We now consider the opposite limit, the kinetic regime where 
$\Gamma_0<<1$.  We can make a perturbative expansion the Bloch function in terms of the  plane  waves, 
$\psi_{0k}=\sum_{m} c_{m}(k)e^{i(k+m2\pi/a)x},$
where $c_{0}(k)=1$ and 
$c_m= U_{m}/[k^2-(k+2\pi m/a]^2)$ 
for $m\ne 0$, with 
$U_m=\int U(x,t)e^{-im 2\pi x/a} dx$ being the Fourier
 coefficient of the effective potential. Because we are interested in 
the neighborhood of $k=0$, no degeneracy needs to be considered. 
 The average velocity  is calculated as 
$J(t)= \frac{2}{i(2\pi/a)^5}(U_{-1}\partial_t U_{1}
-U_{1}\partial_t U_{-1})=\frac{4}{(2\pi/a)^5}|U_1|^2\partial_t \theta_{1},$
where $\theta_1$ is the phase of $U_1$, and the higher order
 terms with $|m| > 1$ have been dropped.  Therefore, $J(t)$
 is proportional to the 
absolute square of the potential and to the rate of phase change in the
 potential. 

To summarize, we have considered a  ratchet model  in which
an overdamped particle is adiabatically
driven by an asymmetric potential which varies periodically both
in space and in  time.
The problem is treated by  transforming the time-dependent 
Fokker-Planck   operator to a time-dependent  hermitian operator. 
The spatial periodicity validates the use of 
adiabatic  theorem, which  allows us to develop an adiabatic 
perturbation theory based on   the 
instantaneous eigenfunctions of the  hermitian operator. 
There are two limiting regimes,
 determined only by which of the potential energy
and the thermal energy dominates. For each of these limiting regimes,
we have discussed  the adiabatic condition, for which
the diffusion constant $D$ sets the time scale. 
We then calculate  the ensemble average of the particle velocity, 
which is independent of $D$. 
The necessity of parity 
symmetry breaking is reflected in that otherwise the relevant integrals 
vanish.   An analytical expression of the average velocity   is 
obtained in terms of the Berry curvature and Berry phase of the Bloch states.
In the deterministic or tight-binding 
limit, i.e. when the potential dominates the thermal energy,
the time average of the average velocity 
is found to be quantized as a Chern number, with
exponentially small  corrections. 
This  analytical  result 
confirms and extends  a previous numerical observation,
and   discloses  the analogy
with the problem of  quantum adiabatic transport. 
On the other hand it provides,
in a  classical statistical system,  
a  physical example  of the 
first Chern   characteristic class, which has been  important   in 
gauge field theory~\cite{yang} and in quantum condensed matter 
physics~\cite{thouless,nt,niu,thouless1}.  We have also discussed the 
kinetic limit, in which the thermal energy dominates the potential. 
Throughout our work,  we have emphasized the effects of
geometry and phases.

We are  indebted to Ben Simons for suggesting 
the analogy between  ratchet 
and quantum adiabatic transport, as well as 
useful discussions. We also thank David Thouless and  
Dima Khemelnitskii for useful discussions. Y.S.  thanks
Peter Littlewood for hospitality.  Q.N. 
was supported by the Welch Foundation, the Texas Advanced Research Program, 
and the US Department of Energy under Grant No. DE-FG03-02ER45958.

\newpage
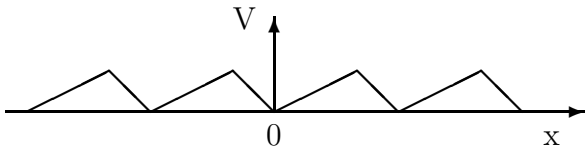
\begin{figure}
\begin{center}
\setlength{\unitlength}{0.55cm}
\begin{picture}(15,3.5)
\thicklines
 \put(-6.5,0){\vector(1,0){14}}
 \put(0,0){\vector(0,1){2.3}}
 \put(-6,0){\line(2,1){2}}
 \put(-4,1){\line(1,-1){1}}
 \put(-3,0){\line(2,1){2}}
 \put(-1,1){\line(1,-1){1}}
 \put(0,0){\line(2,1){2}}
 \put(2,1){\line(1,-1){1}}
 \put(3,0){\line(2,1){2}}
 \put(5,1){\line(1,-1){1}}
 \put(6.5,-0.8){x}
 \put(-1,2){V}
 \put(-0.2,-0.8){0}
\end{picture}
\end{center}
\vspace{1cm}
\caption{An illustration  of a ratchet potential $V(x,t)$
which is  periodic, but asymmetric, in space. In our consideration, it 
is also periodic in time. }
\end{figure}


\begin{references}
\bibitem[(*)]{s} E-mail: ys219@phy.cam.ac.uk
\bibitem[(**)]{n} E-mail: niu@physics.utexas.edu
\bibitem{reimann} For a recent comprehensive review, 
see  P. Reimann, cond-mat/0010237.
\bibitem{prost}  
A. Ajdari and J. Prost, C. R. Acad. Sci. Paris, {\bf 315}, 1635 (1992).
\bibitem{magnasco}   M. Magnasco, Phys. Rev. Lett. 71 (1993) 1477.
\bibitem{feynman} R. Feynman, R.B. Leighton, and M. Sands, {\em The Feynman 
Lectures on Physics} (Addison-Wesley, Reading, MA, 1963), Vol. 1, Chap. 46. 
\bibitem{bartussek} R. Bartussek, P. H\"{a}nggi and J. G. Kissner,
Europhys. Lett. {\bf 28}, 459 (1994). 
\bibitem{parrondo} J. M. R. Parrondo, Phys.~Rev.~E {\bf 57}, 7297 (1998).
\bibitem{chern} 
S. S. Chern, Ann. Math. {\bf 47}, 85 (1946);
see, for example, Y. Choquet-Bruhat {\it et al.},
{\em Analysis, Manifolds and Physics} (North-Holland, Amsterdam, 1983).
\bibitem{thouless} D. J. Thouless, Phys. Rev. B {\bf 27}, 683 (1983).
\bibitem{nt} Q. Niu and D. J. Thouless, J. Phys. A {\bf 17}, 2453 (1984).
\bibitem{niu} Q. Niu, Mod. Phys. Lett. B {\bf 5}, 923 (1991).
\bibitem{zinn} H. Risken, {\em The Fokker-Planck Equation} 
(Springer-Verlag, Berlin, 1989).
\bibitem{kato} T. Kato, J. Phys. Soc. Jap. {\bf 5}, 435 (1950);
A. Messiah, {\em Quantum Mechanics, Vol. II} (North-Holland, Amsterdam, 1961).
\bibitem{yang} T. T. Wu and C. N. Yang, Phys. Rev. D {\bf 12}, 3845 (1975).
\bibitem{thouless1} D. J. Thouless, M. Kohmoto, M. P. Nightingale,
and M. den Nijs,  Phys. Rev. Lett. {\bf 49}, 405 (1982);
 Q. Niu, D. J. Thouless and Y. S. Wu,
 Phys. Rev. B {\bf 31}, 3372 (1985).
\end{references}
\end{document}